\documentclass[sigconf]{acmart}
\usepackage{hyperref}

\PassOptionsToPackage{hyphens}{url}

\usepackage{xspace}
\usepackage{graphicx}
\usepackage{enumitem}
\usepackage{booktabs}
\usepackage{multirow}
\usepackage[binary-units=true]{siunitx}
\usepackage{amsfonts,amsthm,amsmath}
\usepackage{subcaption}
\usepackage{xcolor}
\usepackage{siunitx}
\usepackage{framed}
\usepackage{bm}
\usepackage[ruled,noend]{algorithm2e}
\usepackage{algpseudocode}
\usepackage{tabularx}

\newcolumntype{L}{>{\arraybackslash}X} 
\usepackage[all]{xy}
\DeclareMathAlphabet{\mathsc}{OT1}{cmr}{m}{sc}
\definecolor{chicagomaroon}{rgb}{0.5, 0.0, 0.0}

\iffalse
\newcommand\sarah[1]{\textcolor{red}{Sarah: #1}}
\newcommand\kurt[1]{\textcolor{blue}{Kurt: #1}}
\newcommand\alex[1]{\textcolor{chicagomaroon}{AL: #1}}
\newcommand\geoff[1]{\textcolor{blue}{Geoff: #1}}
\newcommand\todo[1]{\textcolor{blue}{#1}}
\else
\newcommand{\alex}[1]{}
\newcommand{\kurt}[1]{}
\newcommand{\geoff}[1]{}
\newcommand{\sarah}[1]{}
\newcommand\todo[1]{}
\fi

\renewcommand{\paragraph}[1]{\vspace{5pt}\noindent\textbf{#1.}\xspace}

\newcommand*\dash{\ifvmode\quitvmode\else\unskip\kern.16667em\fi---%
\hskip.16667em\relax}

    \copyrightyear{2024}
    \acmYear{2024}
    \setcopyright{rightsretained}
    \acmConference[IMC '24] {Proceedings of the 2024 ACM Internet Measurement Conference}{November 4--6, 2024}{Madrid, Spain.}
    \acmBooktitle{Proceedings of the 2024 ACM Internet Measurement Conference (IMC '24), November 4--6, 2024, Madrid, Spain}
    \acmISBN{979-8-4007-0592-2/24/11} 
    \acmDOI{10.1145/3646547.3689005}

\ccsdesc[500]{Security and privacy~Social network security and privacy}

\keywords{Giveaway Scams; Blockchain Security; Internet Security}

\newcommand{\eg}{e.g., }

\newcommand{\ie}{i.e., }
\newcommand{\etal}{et al.\@\xspace}

\begin{document}

\title{Give and Take: An End-To-End Investigation of Giveaway Scam Conversion Rates}

\newcommand*{\chainanlysis}{\includegraphics[scale=0.020]{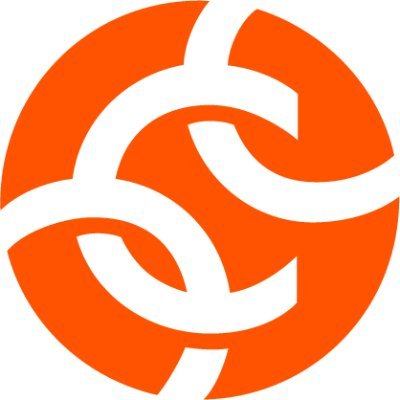}}%

\newcommand*{\google}{\includegraphics[scale=0.015]{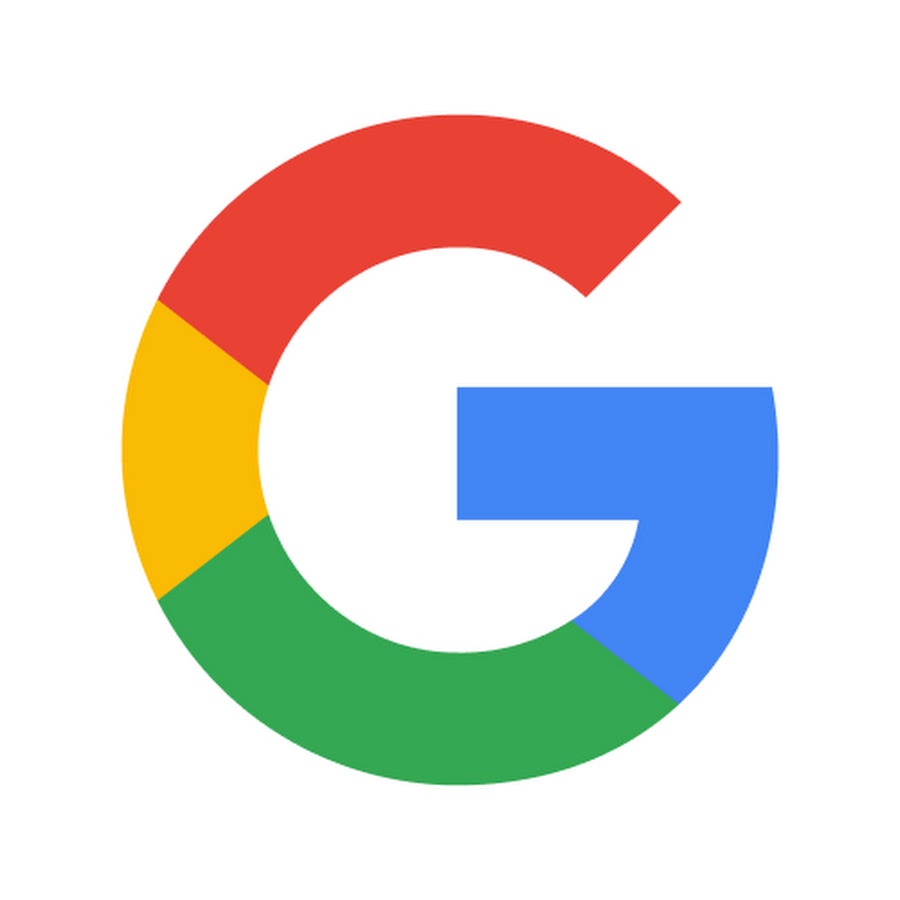}}%

\newcommand*{\ucsd}{\includegraphics[scale=0.020]{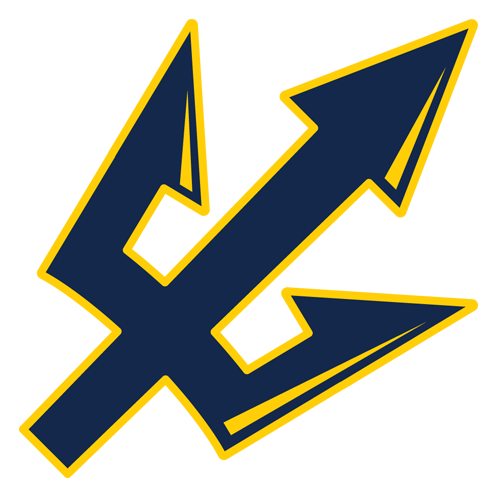}}%

\author{Enze Liu}
\orcid{0000-0003-4288-8485}
\affiliation{%
  \institution{UC San Diego}
\country{}
}

\email{e7liu@ucsd.edu}


\author{George Kappos}
\affiliation{%
  \institution{Chainalysis}
\country{}
}

\email{g.kappos@ucl.ac.uk}

\author{Eric Mugnier}
\affiliation{%
  \institution{UC San Diego}
\country{}
}
\email{emugnier@ucsd.edu}

\author{Luca Invernizzi}
\affiliation{%
  \institution{Google}
\country{}
}
\email{invernizzi@google.com}

\author{Stefan Savage}
\affiliation{%
  \institution{UC San Diego}
\country{}
}
\email{savage@cs.ucsd.edu}

\author{David Tao}
\affiliation{%
  \institution{Google}
\country{}
}
\email{dtao@google.com}

\author{Kurt Thomas}
\affiliation{%
  \institution{Google}
\country{}
}
\email{kurtthomas@google.com}

\author{Geoffrey M. Voelker}
\affiliation{%
  \institution{UC San Diego}
  \country{}
}
\email{voelker@cs.ucsd.edu}

\author{Sarah Meiklejohn}
\affiliation{%
  \institution{Google}
  \country{}
}
\email{s.meiklejohn@ucl.ac.uk}





\renewcommand{\shortauthors}{Enze Liu et al.}
\renewcommand{\authors}{Enze Liu, George Kappos, Eric Mugnier, Luca Invernizzi, Stefan Savage, David Tao, Kurt Thomas, Geoffrey M. Voelker, Sarah Meiklejohn}



%
%



\begin{abstract}

The Internet's combination of low communication cost, global reach, and functional anonymity has allowed fraudulent scam volumes to reach new heights.
Designing effective interventions 
requires first understanding the context: 
how scammers reach potential victims, the earnings they make, and any potential bottlenecks for durable interventions. In this short paper, we focus on these questions in the context of cryptocurrency giveaway scams, where victims are tricked into irreversibly transferring funds to scammers under the pretense of even greater returns.
Combining data from Twitter,\footnote{The platform has been known as X since July 2023, but most of our analysis pertains to tweets posted before then so we continue to call it Twitter.} YouTube and Twitch livestreams, landing pages, and cryptocurrency blockchains, we measure how giveaway scams operate at scale. We find that 1 in 1000 scam tweets, and 4 in 100,000 livestream views, net a victim, and that scammers managed to extract nearly \$4.62 million from just hundreds of victims during our measurement window.

\end{abstract}

\maketitle





\section{Introduction}

Since the introduction of the Bitcoin whitepaper in 2008, cryptocurrencies and blockchain-based technologies have attracted attention from investors, technologists, researchers, governments, and the general public.  Unsurprisingly, this rapid rise in popularity has also attracted scammers seeking to exploit the gap between people's (high) interest in these technologies and their (low) technical understanding of their operation.  Today, there are many types of scams run within the cryptocurrency ecosystem: Ponzi and pyramid schemes~\cite{vasek2015freelunch,vasek2018ponzi,bartoletti2018ponzi,chen2018ponzi,siu2022bitcointalk,kell2023forsage}; NFT ``rug pulls''~\cite{roy2023nftscams,huang2023rugpulls}, counterfeiting~\cite{das2022nft},  airdrop scams~\cite{roy2023nftscams}; pump-and-dump schemes~\cite{dhawan2023pumpdump,hamrick2018pumpdump,kamps2018pumpdump,xia2020covidscams,xu2019pumpdump};  
and giveaway scams~\cite{vakilinia2022giveaway,li2023giveaway,li2023twitter-giveaway}.

In a giveaway scam, users are enticed to visit a website that promises to give away a certain number of (cryptocurrency) coins. Often, the website leverages branding associated with a prominent public figure or company to increase its perceived legitimacy.  To participate, users are instructed to send some coins to a cryptocurrency address specified on the website and are promised they will receive double the amount in return.  These promised funds fail to  materialize and, given the irreversible nature of cryptocurrency transactions, there is no way for the user to recover their money.

The existence of giveaway scam sites is well documented both in recent academic literature~\cite{vakilinia2022giveaway,li2023twitter-giveaway,li2023giveaway} and by online tech journalists~\cite{toulas23,ahmed23}.  However, our understanding of how users are lured to these sites, the extent of victim participation, and the conversion rate that scammers can expect 
is largely anecdotal.  Indeed, while there is received wisdom that scammers utilize mainstream ``broadcast'' style media such as Twitter, the extent to which such platforms represent the dominant source of such activity is unclear.  Further complicating analysis is the fact that livestreaming platforms (e.g., YouTube and Twitch) often do not provide a retrospective record and thus must be analyzed contemporaneously with their use.  

In this paper, we present an empirical analysis of cryptocurrency giveaway scams, conversion rates, and overall revenue.  We focus on scams appearing on Twitter, YouTube, and Twitch, which Li et al.\ suggest are the most prevalent platforms for such activity~\cite{li2023giveaway}.  From these platforms we collect posts (retrospectively) and streams (prospectively) and identify and crawl both existing and new promoted
scam giveaway sites. 
From these, we extract
the cryptocurrency addresses used to solicit victims for funds and quantify the revenue of these scams; we also explore different aspects of victim behavior (e.g., how victims pay in) and scammer behavior (e.g., how scammers promote on Twitter and YouTube as well as cash out). Together this data provides a broader vantage point for understanding the revenue model of giveaway scams and potential paths to interventions.


\section{Related Work}
Our work builds on a number of prior efforts that have investigated cryptocurrency giveaway scams. Xia et al.~\cite{xia2020covidscams} document dozens of giveaway scams while searching for Covid-themed cryptocurrency fraud reported by a variety of industry and government sources. Phillips and Wilder~\cite{phillips2020tracing} analyzed hundreds of giveaways scams (which they refer to as advanced-fee scams) crowdsourced by victims and researchers on CryptoScamDB and URLScan.  Li et al.~\cite{li2023giveaway} deployed the CryptoScamTracker tool, which found thousands of scam domains by examining Certificate Transparency (CT) data for a six-month period between January and June 2022.  The authors used a combination of scam-related keywords to identify likely scam domains, filtering down this list and then validating manually (i.e., that the site does host a scam) to obtain their scam domains.  

These efforts generally focus on scam sites themselves, however, and do not capture the modality by which victims are lured to visit.  One recent such study is Vakilinia's qualitative analysis of giveaway scams livestreamed over two days on YouTube~\cite{vakilinia2022giveaway}. While small in scale, this work helps confirm our understanding of livestreamed scams and thus our methodology for finding them.  Another example is Li et al.'s contemporaneous paper describing GiveawayScamHunter, which explores giveaway scams specifically promoted via Twitter Lists~\cite{li2023twitter-giveaway}.  

In addition to giveaway scams, the security community has explored a range of other cryptocurrency-related scams, including high-yield investment scams~\cite{vasek2015freelunch, bartoletti2018ponzi, vasek2018ponzi, boshmaf2020investigating, chen2019exploiting, jung2019data, toyoda2019novel, chen2018ponzi, siu2023get, bartoletti2020dissecting, kell2023forsage, siu2022bitcointalk}; 
fake or high-risk cryptocurrency-related services (e.g., arbitrage bots~\cite{li2023towards},  exchanges~\cite{vasek2015freelunch, moore2013beware, xia2020characterizing}, 
pump-and-dump scams~\cite{dhawan2023pumpdump,hamrick2018pumpdump,kamps2018pumpdump,xia2020covidscams,xu2019pumpdump}; and
new coins and tokens~\cite{gao2020tracking, xia2021trade}); 
initial coin offering scams~\cite{sapkota2020much, zetzsche2017ico, liebau2019crypto, tiwari2020future, bian2018icorating};
sextortion scams~\cite{paquet2019spams, oggier2020ego}; technical support scams~\cite{acharya2024conning}; 
scams and fraud in non-fungible token markets~\cite{kshetri2022scams, das2022nft};
and cryptocurrency-themed phishing websites~\cite{andryukhin2019phishing, holub2018coinhoarder, phillips2020tracing, he2023txphishscope}.  
In contrast to scams that require more sophisticated methods for
turning a profit --- price manipulation (e.g., pump-and-dump),
hierarchies of participants to asymmetrically channel funds (e.g.,
Ponzi schemes), etc. --- giveaway scams are quite straightforward.
Victims simply give their funds to scammers.

Finally, another relevant area of research focuses on understanding the risk models and security practices of cryptocurrency users~\cite{abramova2021bits, alshamsi2019user, guanexamining, ooi2021embracing, frohlich2021don, krombholz2017other}. These studies help shed light on why or how users fall for scams in general, but do not specifically target giveaways or similar scams.

\section{Methodology}
Here, we describe our various sources of data and
explain our methodology for identifying giveaway scams on social media (Twitter) and livestreams
(Twitch and YouTube).




\subsection{Data sources}
\label{sec:twitter-collection}
We rely on three pre-existing datasets for our analysis: a snapshot of Twitter, a corpus of known giveaway scam domains and cryptocurrency addresses, 
and a snapshot of blockchain transactions. We add to this a new dataset derived from monitoring livestreams of giveaway scams and crawling the leads presented via the streams (discussed in Section~\ref{sec:method-youtube}).

\paragraph{Twitter dataset}
Our Twitter dataset originates from Google's Internet-wide crawl of public URLs, which respects robots.txt and other rules for crawlers. This means our analysis considers only publicly available tweets. Our dataset covers a period from April 1, 2020 until August 20, 2023. It does not contain every single tweet, and is biased towards tweets that are more discoverable (e.g., posted by accounts with a non-zero number of followers).  Within Google, access to this data was granted to us only due to the nature of our research project and its potential to disrupt abusive behavior. In total, our snapshot contains tens of billions of tweets posted by hundreds of millions of accounts.

\begin{figure}[t]
\centering
\begin{subfigure}[b]{0.20\textwidth}
\includegraphics[width=\textwidth]{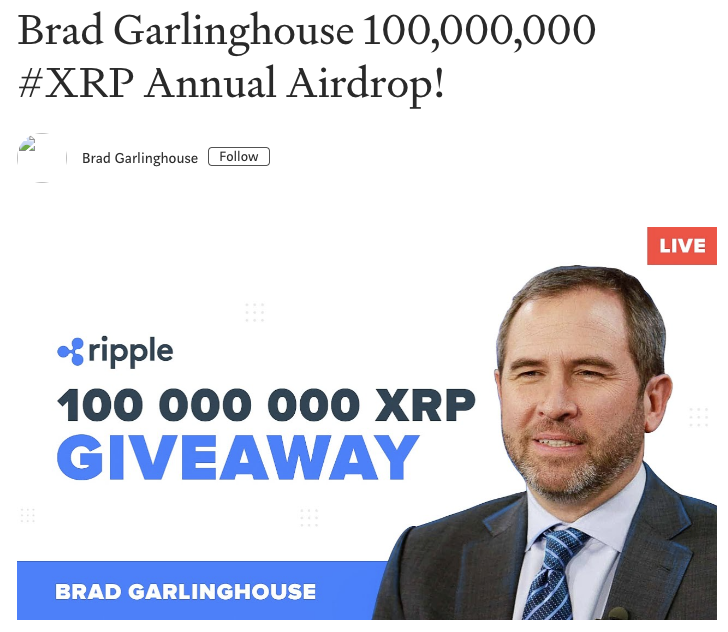}
\end{subfigure}
\hspace{5mm}
\begin{subfigure}[b]{0.20\textwidth}
\includegraphics[width=\textwidth]{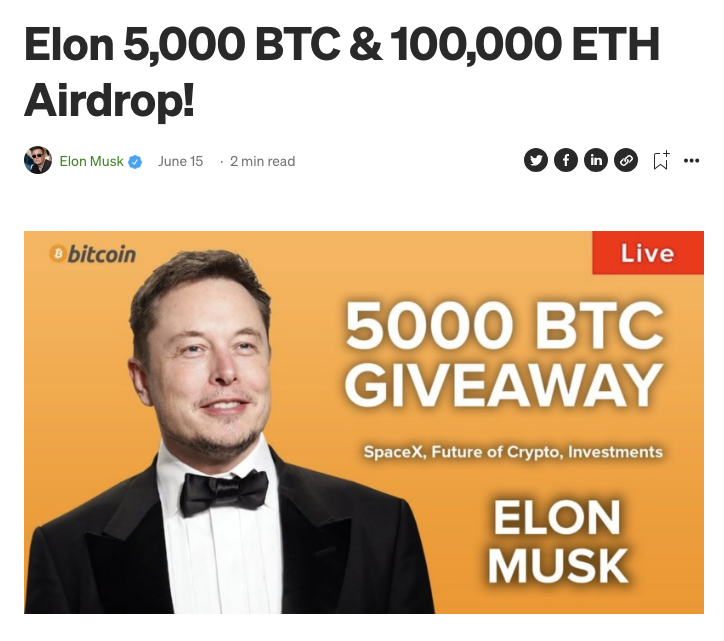}
\end{subfigure}
\caption{Example giveaway scam landing pages promoted via Twitter. Scammers impersonate popular personalities including Brad Garlinghouse (the Ripple CEO) and Elon Musk.\vspace{-0.3cm}}\label{fig:twitter_sample_domains}
\end{figure}

\paragraph{Giveaway scam dataset}
We rely on a manually curated dataset made available by Li \etal~\cite{li2023giveaway} containing 3,863 domains previously identified as cryptocurrency giveaways over a period from January 1, 2022 to June 29, 2022. This dataset annotates each domain with the cryptocurrency addresses published at the time it was crawled. (Websites often use addresses for multiple different cryptocurrencies.) 
%
Starting from this dataset, we identified all tweets that contained at least one known scam domain, totaling 457,248 tweets from 33,841 distinct accounts (Table~\ref{tab:dataset-size}). Of the original 3,863 scam domains, only 361 (9\%) appeared on Twitter---a sample of which we show in Figure~\ref{fig:twitter_sample_domains}--- indicating that many scams are likely promoted through other means.


\paragraph{Blockchain data} We have access to the Chainalysis Crypto Investigations tool~\cite{chainanalysis}, which processes raw blockchain data and annotates each address indicating (1) the \emph{multi-input cluster}~\cite{reid2013bitcoin,meiklejohn2013fistful} it belongs to, meaning the cluster of addresses that certain heuristics indicate are all operated by the same entity, and (2) the likely \emph{category} (\eg exchange, mixer, token smart contract) of the real-world operator of that cluster.\footnote{The tool can be configured to provide the name of the cluster operator, but our agreement with Chainalysis meant we had access only to the broader category.}  These latter annotations are obtained via regular transactions that Chainalysis conducts with known cryptocurrency services, such as exchanges, and the former annotation avoids false positives (such as those induced by Coinjoin transactions~\cite{coinjoin}) using proprietary heuristics. While there is some potential for error in both of these areas, Chainalysis data has been used in previous academic studies~\cite{portnoff2017backpage,huang2018tracking,kappos2022peel} and was recently ruled to be sufficiently reliable as to be admissible evidence in court~\cite{chainalysis-daubert}.

\subsection{Identifying giveaway livestreams}
\label{sec:method-youtube}
\label{sec:livestream-pipeline}

\begin{figure}[t]
\centering
\includegraphics[width=\columnwidth]{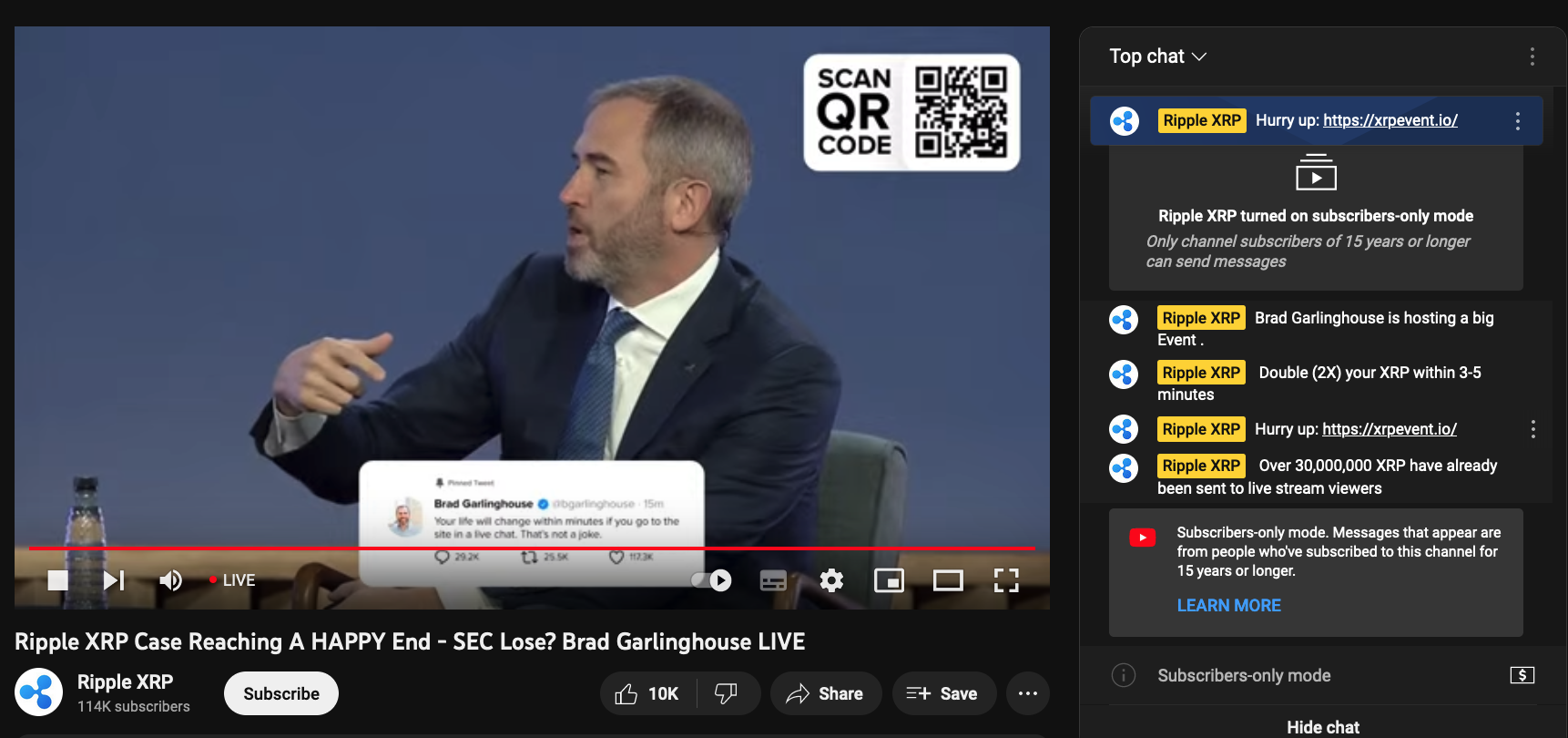}
\caption{Example livestream containing a giveaway scam. The video playing is of Brad Garlinghouse 
and the scam website is linked to in both the chat and the embedded QR code.\vspace{-0.3cm}}
\label{fig:sample-livestream}
\end{figure}

We built a measurement pipeline for identifying and capturing giveaway scams on the YouTube and Twitch streaming services. These scam livestreams typically play a pre-recorded video that ultimately directs viewers to a scam page via a link or QR code. Figure~\ref{fig:sample-livestream} shows such an example, in which the scam page is promoted both as a QR code and as a link in the chat.

Our pipeline polled available livestreams via platform APIs, recorded them, crawled the pages promoted by the QR codes and links, and annotated each scam domain with 
blockchain metadata. The design decisions are based on a 14-day pilot study conducted July 1--14, 2023. While we crawled both YouTube and Twitch, we did not detect any giveaway scams during our collection window for Twitch despite past reports of such scams~\cite{twitch-scams2022}. As a result, we focus our subsequent discussion and analysis on YouTube, deferring Twitch-specific details to Appendix~\ref{subsec:twitch_details}.

\paragraph{Identifying potential giveaway streams} 
Our pipeline starts data collection by periodically querying the YouTube API for available streams and associated metadata. Given a set of keywords, the API returns livestreams that are associated with the keywords (in a manner similar to Google Search). 
Our system queried the API at the maximum rate allowed, which meant retrieving the list of livestreams and channel metadata (e.g., number of subscribers) every 30 minutes and retrieving livestream metadata (e.g., concurrent and total viewers) every 7.5 minutes. 

We derived our set of search keywords using a combination of popular cryptocurrency names and those used by Li et al.~\cite{li2023giveaway}.
In more detail, we opted for a set of keywords consisting of the names and ticker symbols of the top 20 cryptocurrencies (as listed on \url{coinmarketcap.com} on July 1, 2023),\footnote{In three cases (for ADA, SOL, and DOT), we appended the word ``coin'' to the ticker symbol to avoid introducing irrelevant streams.} along with the keywords used by CryptoScamTracker~\cite[Table XI]{li2023giveaway} and two catch-all terms: ``cryptocurrency'' and ``crypto.'' We discuss the effectiveness of these keywords in more detail in Appendix~\ref{subsec:effectiveness_of_search_keywords}.
%


\paragraph{Stream and chat recording} 
The system recorded each livestream (using Streamlink~\cite{Streamlink:online}) and its associated chat (using the YouTube API) to identify linked URLs.  Our final pipeline recorded videos for two seconds at a time, with snapshots taken every 7.5 minutes, to capture QR codes embedded in the video.  
For chats, it polled the chat API every 7.5 minutes to obtain the last 70 historical messages (the maximum set by YouTube).

We chose these parameters because in our pilot study we observed that URLs either appeared in chat or
were embedded in the video using a QR code. We thus considered (1) how long a QR code lasts once it appears and (2) how many chat messages are posted during the lifespan of a scam livestream. We started by running our pipeline at maximum capacity on YouTube, with the default setup of recording a stream for two seconds as well as retrieving up to 70 historical chat messages every 7.5 minutes. In addition, for a subset of the livestreams, we kept recording the stream upon detecting a QR code.
With this setup, we manually identified 276 streams that promoted 59 unique giveaway websites. We first examined how long a QR code lasted after we first detected it. For 41 scam livestreams, we continued to record them after detecting a QR code.
We found that for a majority of the livestreams, a QR code remained displayed continuously for a reasonable amount of time (mean of 7,200 seconds and median of 3,140 seconds), except for one case where the QR code was displayed periodically for around 15 seconds. This finding suggests that two-second samples are enough to capture a QR code in the livestream, if present. Second, we examined scam streams that had chat messages. We found that all scam streams had few chat messages (less than 10) and no user interactions, and thus decided that 70 historical chat messages every 7.5 minutes was sufficient.

\paragraph{Extracting URLs and QR code leads} 
From our livestream data, the system identified potential links to scams by (1) extracting URLs from chats via regular expressions; and (2) extracting and interpreting QR codes via visual analysis of the captured video frames using the \texttt{opencv} and \texttt{pyzbar} Python libraries. We use two libraries to increase the detection rate.



\paragraph{Crawling potential scam URLs} 
The system crawled each potential scam URL, revisiting both old and newly discovered websites daily until (1) we reached the end of our collection period, or (2) fetching the URL resulted in an error three days in a row. Prior work~\cite{li2023giveaway} suggested that some giveaway scam sites adopted anti-bot tactics. We identified four types of cloaking behavior in our pilot study: (1) IP-based cloaking, in which requests from an institutional network resulted in 403 responses but ones from a residential network did not; (2) user agent-based cloaking, in which requests from browsers that were not running on Windows or Mac resulted in 403 responses; (3) customized front pages, in which human interaction was required to click on a button or select a desired cryptocurrency; and (4) anti-bot detection through Cloudflare. We counter these behaviors by: (1) using a VPN to mask the system IP address; (2) spoofing the User-Agent string to appear as a popular device and browser; (3) having a heuristic module to click through common front pages; and (4) verifying the crawler with Cloudflare to bypass anti-bot detection.\footnote{\url{https://blog.cloudflare.com/friendly-bots}}


\paragraph{Validating scam URLs and identifying cryptocurrency addresses} 
We verified that every final clickthrough or landing page was related to a giveaway scam by (1) ensuring there was a valid cryptocurrency address published on the site. In addition, we require that either (2) the landing page included a set of heuristic keywords related to scams (\eg ``hurry'' or ``participate'') or (3) the domain contained a common set of scam keywords. For (1), we extracted addresses via a regular expression and then validated the address.\footnote{We used  coinaddrvalidator~\cite{coinaddrvalidator-github:online} and multicoin-address-validator~\cite{multicoin-address-validator-npm:online}, and considered an address valid if either tool indicated that it was valid.} For (2) and (3), we relied on the same keywords from the CryptoScamTracker tool. 
We then manually examined all sites (4,611) that met the requirements. To ensure that our heuristics do not miss a significant number of domains, we randomly sampled 389 sites that had only a valid cryptocurrency address and did not find any additional scam sites.

\paragraph{Final dataset}
\label{sec:livestream-dataset}
We ran our measurement pipeline from July 24, 2023 to January 21, 2024 (26 weeks), identifying a total of 2,069 livestreams from 1,632 different channels (Table~\ref{tab:dataset-size}). These streams linked to 343 distinct domains (all but one domains met criteria 2 and 303 domains met criteria 3).
Our data is continuous throughout this period apart from 11 days of infrastructure outages.\footnote{These dates are: 2023-08-15, 2023-08-16, 2023-09-01, 2023-09-28, 2023-10-06, 2023-11-18, 2023-11-19, 2023-12-26, 2023-12-12, 2024-01-06, and 2024-01-21.}

\begin{table}[t]
\centering
\begin{tabular}{lS[table-format=3.0]S[table-format=6.0]S[table-format=7.0]}
\toprule
\bf Source &\bf  {Domains} &\bf  {Accounts} &\bf  {Artifacts} \\
\midrule
Twitter & 361 & 33841 & 457248 \\
YouTube & 343 & 1632 & 2069 \\
\bottomrule
\end{tabular}
\caption{The number of objects in our collected datasets.  Accounts in the YouTube dataset correspond to channels, and artifacts are individual tweets or livestreams.\vspace{-0.3cm}}
\label{tab:dataset-size}
\vspace{-12pt}
\end{table}

%

\subsection{Limitations}\label{sec:limitations}
As with any measurement study, there are a number of limitations with our methodology. 
Our Twitter dataset may omit some tweets, resulting in us underestimating the volume of scam tweets or the revenue generated by scams; likewise, this may happen due to missing or inaccurate cryptocurrency addresses in the CryptoScamTracker dataset (as indeed we observed manually). 
Given that only 9\% of the domains previously identified by CryptoScamTracker overlap with Twitter, it is also clear that many scams are promoted via other distribution channels---be it social media, email, or messaging. Similarly, our keyword-based matching strategies for YouTube and Twitch likely resulted in omitting some scams. Finally, our Twitter and YouTube data are drawn from non-overlapping periods of time (early 2022 vs.\ late 2023), which means the tactics of scammers may have evolved so we cannot directly compare profitability. Despite these limitations, we are able to draw insights into the promotion of giveaway scams and provide the first estimates on the conversion rate and revenue of such campaigns.
\section{Giveaway Scam Lures}
We begin our analysis by examining the volume of tweets and livestreams (i.e., ``lures'') that ultimately directed victims to scam landing pages; the tactics used by scammers to make their lures discoverable; and which cryptocurrencies were targeted by scammers.



\begin{figure}[t]
\centering
\vspace*{-0.1in}
\includegraphics[width=0.9\columnwidth]{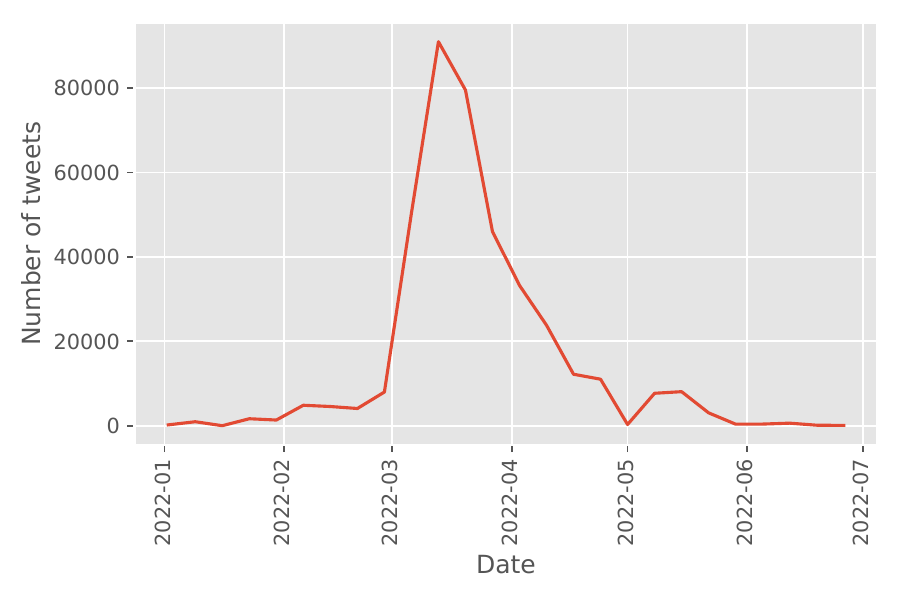}
\vspace{-0.5cm}
\caption{The volume of giveaway scam tweets on a given week, between January 1, 2022 and July 7, 2022.\vspace{-0.3cm}}  
\label{fig:tweets-days}
\end{figure}


\subsection{Volume} 
\begin{figure}[t]
\centering
\includegraphics[width=0.8\columnwidth]{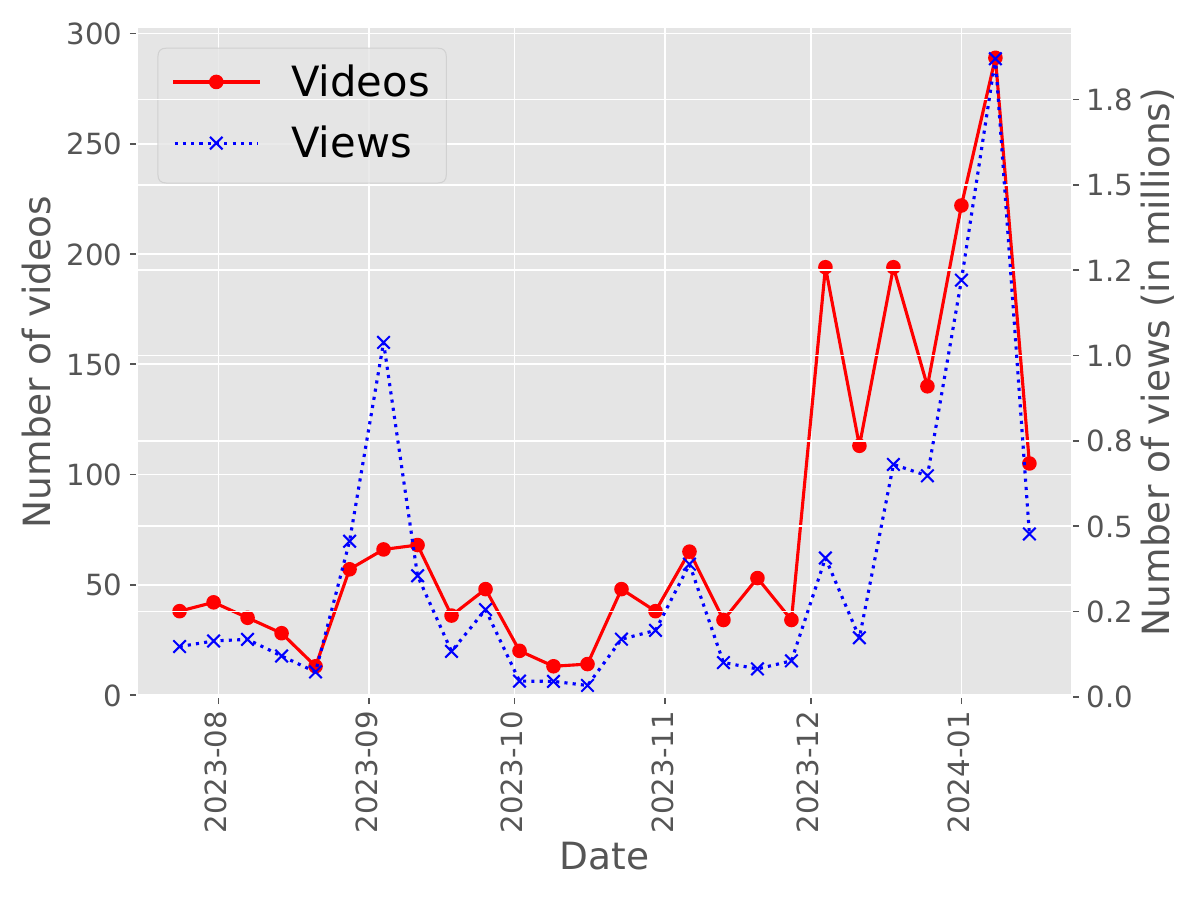}
\vspace{-0.5cm}
\caption{The volume of giveaway scam streams and views per week, between July 24, 2023 and January 21, 2024.\vspace{-0.4cm}} 
\label{fig:ls-domains-per-day}
\end{figure}

In total, we observed 457,248 scam tweets and 2,069 scam livestreams. We present a weekly timeline of scam activity for Twitter in Figure~\ref{fig:tweets-days} and YouTube in Figure~\ref{fig:ls-domains-per-day}. While scam lures appeared consistently throughout our measurement window, there are notable bursts of activity. Twitter had a single peak in activity in March 2022, with scammers posting a maximum of 90,984 tweets in a single week. YouTube had a burst in September 2023 and again over the holiday period of Dec--Jan 2024, peaking at 289 streams and 1,869,399 views in a single week. While these scams are smaller in scale compared to spam email campaigns involving tens of millions of messages~\cite{kanich2008spamalytics}, as we show shortly, they are nevertheless highly profitable.

%
%

\subsection{Discoverability}
\label{sec:visibility}
To reach a broad audience on Twitter, scammers hashtagged their tweets 96\% of the time (\eg using a cryptocurrency's name or ticker symbol). 
Users searching for the latest news related to popular coins (\eg \texttt{\#btc}, \texttt{\#eth}) might then interact with the scammer's tweets. While scammers also pursued alternative 
tactics on Twitter, such as mentioning users or replying to popular tweets, these strategies occurred in only 0.1\% and 0.3\% of tweets respectively.\footnote{
For Twitter, our dataset lacks any information on the number of views per tweet, or the following or follower counts of accounts. As such, we focus on discoverability.} For YouTube, scammers reached victims using a combination of organic viewership and search discoverability. Channels used by scammers to host livestreams had a median of 16.8K subscribers, with the single largest channel having 19 million subscribers---likely a compromised pre-existing channel. Beyond existing audiences, 93\% of livestreams included a cryptocurrency keyword in their title, their description, or the name of the associated channel.

\subsection{Currencies}
\label{sec:popular-currencies}
To identify the cryptocurrencies most commonly targeted by scammers, we searched for relevant keywords in each tweet's hashtags and each livestream's title, channel name, and description, using the names and ticker symbols of the top 20 coins with the highest market capitalization (\eg \texttt{Bitcoin} and \texttt{btc}).\footnote{We fetched the coins and their ranking from \url{coinmarketcap.com} on August 7, 2023.}   
For Twitter, of these top 20 coins, Ripple was by far the most popular choice, matching 91\% 
    of the tweets in our dataset, followed by Ethereum (12\%) and Bitcoin (7\%). For YouTube, Bitcoin was the most popular coin (matching 65\% of the livestreams), followed by Ethereum (49\%) and Ripple (40\%). 
Totals do not add to 100\% as tweets and livestreams could reference multiple coins. The difference in currencies may be due to changes in tactics by scammers, with at least a year transpiring between our two datasets. 

\section{Giveaway Scam Payments}
\label{sec:profit}
Using our raw blockchain data, we retrospectively analyzed payments to scammers to explore the revenue they reaped, the conversion rate per lure, the behavior of victims, and the payment infrastructure relied upon by scammers.

\subsection{Isolating relevant payments}
We focused on the three most popular coins targeted by scammers (per Section~\ref{sec:popular-currencies}): Bitcoin (BTC), Ethereum (ETH), and Ripple (XRP). For each coin, we identified all incoming transactions to cryptocurrency addresses that appeared on a giveaway scam landing page posted to Twitter or YouTube. 
We filtered the set of incoming transactions to an identified scam address in two key ways.  First, following Gomez et al.~\cite{gomez2023cybercrime}, we ignored transactions where the sender is a known scam address since this represents the consolidation of funds rather than a payment from an actual victim.  Second, we focused on payments that co-occurred with a lure to capture the lure's effectiveness in driving victims to the scam.  We acknowledge that the resulting set of payments might both include non-victim transactions (from unknown scammer addresses) and exclude victim transactions (due to lures not in our data); nevertheless, we use it as an approximation of victim payments to the scam.  We believe this estimate is more accurate than previous studies, which considered all incoming transactions~\cite{li2023giveaway}, but also compare with the total revenue (across all transactions) and leave as future work refinements to our analysis that take other behaviors into account. 

\subsection{Twitter scam revenue}
\label{sec:twitter-blockchain}
  
The 361 domains promoted by scammers on Twitter used 186 distinct addresses (with some domains having multiple addresses due to accepting different cryptocurrencies). After filtering out domains that did not use any BTC, ETH, or XRP addresses, we were left with 258 domains. Of these, only 121 (47\%) received any incoming transaction; \ie fewer than half of the domains received any transaction at all. For the 121 domains that received a transaction, we compared the transaction times to the times of the tweets promoting the scam domain. We considered a payment as co-occurring if it happened within one week of a tweet's appearance.  Despite this generous window, only 43\% of payments (695 of 1,633) fell within it.  Of these, we removed a further 24 payments that came from known scam addresses, resulting in a final dataset of 671 payments.  
 
Table~\ref{table:profit} summarizes the overall co-occurring revenue made across
Twitter and YouTube, normalizing payment values across cryptocurrencies
using the average USD price of each coin on the day of the payment.\footnote{We used historical data from \url{https://finance.yahoo.com/crypto/}.}  In total, we estimate that Twitter-based giveaway scams
yielded \$2.7M in revenue. 
If we include all incoming transactions to these addresses then the total jumps to \$6.6M; as suggested above, the real revenue likely lies between the two numbers.

\begin{table}[t]
\centering
\begin{tabularx}{\columnwidth}{XS[table-format=7.0]S[table-format=6.0]}
\toprule
\bf Metric & \bf {Twitter} & \bf {YouTube} \\
\midrule
Payments (co-occurring)  & 671 & 638 \\
Payments (any)  &  1633 & 2074 \\
\midrule
USD revenue (co-occurring) & 2693009 & 1932654 \\
\hspace{5pt}\textit{from BTC}  & 1269579 & 1422065 \\ 
\hspace{5pt}\textit{from ETH}  & 442583 & 266693 \\ 
\hspace{5pt}\textit{from XRP}  & 980847 & 243896 \\
\midrule
USD revenue (any)  & 6598691 & 4705978 \\
\bottomrule
\end{tabularx}

\caption{Revenue of giveaway scams on Twitter and YouTube livestreams, considering only co-occurring payments made in BTC, ETH, or XRP. For completeness, we also include all payments to the addresses, irrespective of what might explain the payment origin.\vspace{-0.8cm} 
}
\label{table:profit}
\end{table}


\subsection{YouTube livestream scam revenue}

Across the 343 domains promoted via YouTube, 
342 domains had a BTC, ETH, or XRP address.
Of these domains, 231 (67\%) had addresses that received at least one transaction. As with Twitter, we looked at the timing of each incoming transaction
for livestream addresses.  We considered it as co-occurring with a
livestream if the transaction occurred during the
stream or up to eight hours after it ended. Our goal was again to provide a generous window, but the result was similar: only 34\% of
payments to the addresses (695 of 2074) fell within this time interval.  We removed another 57 payments from known scam addresses, resulting in our final dataset of 638 payments.



Table~\ref{table:profit} summarizes the overall revenue made by
livestream scams.  In total, we estimate that livestream-based
giveaway scams yielded \$1.9M. If we include all incoming transactions to these addresses, the total increases to \$4.7M.


\subsection{Victim behavior}
\label{sec:victim-behavior}



\paragraph{Conversions}
The 671 Twitter-related payments had 528 unique senders, and the 
638 livestream-related payments had
399 unique senders.  
%
These metrics put into context the huge volume of lures necessary to attract a victim that ultimately pays out. For Twitter, the conversion rate of tweets (with associated cryptocurrency addresses) to victims was 0.12\%---or roughly 1 in 1000 tweets netting a victim. For YouTube, the conversion rate of viewers to victims was 0.0039\%---or roughly 4 in 100,000 views netting a victim. We caution comparing these figures directly as there is an unknown number of viewers per tweet. 



\paragraph{Payment origins} 
For the 1,309 payments across Twitter and YouTube, 755 (58\%) came from centralized exchanges;\footnote{As we saw only the category of each sender, rather than the name, we cannot break down such senders into individual exchanges.} the results of Phillips and Wilder suggest that this number would increase if we used more advanced blockchain analysis to track the source of funds~\cite{phillips2020tracing}.  The prevalence of exchanges is not surprising, as we would expect many victims of this type of scam to have a low level of comfort making cryptocurrency transactions directly.


\paragraph{Payment distribution}  
The total revenue was dominated by a small number of transactions: the top 24 Twitter-related payments captured 50\% of the value, while the top 164 payments captured 90\% of the value.  YouTube scams exhibited a similar phenomenon: 20 payments captured 50\% of the value and 147 payments 90\% of the value.
These results suggest that the giveaway scam revenue model is
significantly influenced by large transactions, where the goal for
promoting the scam is to cast a wide net and, in doing so, capture a
small fraction of victims who will give large amounts (analogous to
the whale profit model for free-to-play mobile
games~\cite{udonis-mobile-whales24}).  Such a model notably differs
from email spam, in which the variance in
individual customer payments is moderate and leads to a business model
driven by order volume from new and repeat customers~\cite{pharmleaks:usesec12}.

\subsection{Scammer behavior}

Across the 1,309 payments, there were only 339 distinct recipients: 68
for Twitter-related payments and 271 for YouTube-related ones; given the similar number of payments, this suggests scammers running YouTube campaigns made more of an effort to cycle between different addresses.  Of these, 145 of 166 (87\%) of the BTC addresses were in a multi-input cluster of size one, which 
suggests that scammers made an explicit effort to
prevent address clustering.

Similarly, when looking at outgoing transactions from these addresses, only 4\% of the recipients (57 of 1,363) were labeled as belonging to a centralized exchange.  Another 87\% of recipients were unlabeled, which again suggests the need for more advanced blockchain analysis~\cite{phillips2020tracing} to identify additional exchange interactions.
Of the remaining categories associated with recipients, some suggested meaningful knowledge of how to use
cryptocurrency (e.g., 13 recipients were tagged as `token smart
contract' and four as `mixing') 
and others suggested that the
scammers operated within a larger illicit ecosystem (22 recipients were tagged as `scam' and 13 as `sanctioned entity').

Altogether, these results suggest a profitable business for a set of knowledgeable and technically sophisticated scammers who take advantage of a set of victims with low technical sophistication and knowledge about cryptocurrencies.

\section{Discussion}\label{sec:discussion}
Based on our findings, we compare the revenue of giveaway scams relative to past scams and discuss the prospects of interventions to undermine giveaway scams.

\subsection{Approximating scam revenue} 
Previous studies of pharmaceutical spam estimated a conversion rate of 0.00001\%~\cite{kanich2008spamalytics}.
While our conversion rates are not directly comparable (\eg tweets or livestreams vs.\ emails), they are markedly higher than email-based spam: 0.12\% of tweets and 0.0039\% of livestream views yielded a victim. We again acknowledge that the conversion rate for Twitter is based on the number of tweets rather than the number of views (as the number of views per tweet is unknown), the set of payments might include irrelevant or exclude relevant transactions (thus artificially inflating or deflating the derived rate), and 
the analyses take place in different time periods.
Nevertheless, these factors are unlikely to account for the orders-of-magnitude difference with email spam. 

Multiple factors may explain the higher returns for cryptocurrency giveaway scams. First, the relatively nascent abuse landscape of cryptocurrency may mean users are less familiar with fraud and its risks as compared to email spam. Second, the rags to riches narratives surrounding cryptocurrency may make a headline of ``5000 BTC giveaway'' more believable, particularly when there are real headlines such as ``Elon Musk will give away 1 million Dogecoin if you can prove his family own emerald mine''~\cite{elon-giveaway}.



\subsection{Interventions} 
A natural question for any scam is where bottlenecks exist that might be ripe for disruption. 
Here, one choice appears to be the payment origins of victims, as at least 58\% of victims relied on centralized exchanges. Early detection of payments to scammers via these exchanges represents a sizable and durable protection~\cite{metamask}: scammers likely cannot influence the exchange that victims use, as we suspect most victims have low comfort with using cryptocurrency directly. In addition, similar to what banks already implement, exchanges could implement
additional warnings when users attempt to make large transactions, which constitute a large portion of the revenue in our study. That said, detecting the cryptocurrency addresses engaging in fraud and informing these exchanges promptly remain challenging problems and require collaboration across multiple sectors. %









\section{Conclusions}
In this paper, we provided a first look at cryptocurrency giveaway scams conducted via social media and livestreams. 
Scammers collectively earned \$2.7M via Twitter scams and \$1.9M via livestreams. These earnings reflect a conversion rate of 0.12\% per tweet and 0.0039\% per livestream viewer.  
We believe that interventions require a defense-in-depth approach. Apart from traditional account-based and blocklist-based security measures, cryptocurrency exchanges appear to be a promising bottleneck for depriving scammers of the means to transact with victims.


\section*{Acknowledgements}
We thank our anonymous shepherd and reviewers for their insightful suggestions and feedback. 
We also thank Xigao Li for his advice on setting up the crawling infrastructure, Babak (Bob) Amin Azad for helping us verify our bot with Cloudflare,
and Jennifer Folkestad and Cindy Moore for operational support.
This work was supported in part by National Science Foundation grant
CNS-2152644.


{\footnotesize
\bibliographystyle{abbrv}
\interlinepenalty=10000

}
 
\appendix
\begin{table*}[t]
  \begin{tabularx}{\textwidth}{lL}
    \toprule
    \textbf{Category} & \textbf{Keywords Used}\\ 
    \midrule
    Coin names & bitcoin btc ethereum eth tether usdt ripple xrp bnb ``usd coin'' usdc cardano ada coin ada dogecoin doge solana sol tron trx litecoin ltc polkadot dot polygon matic wrapped bitcoin wbtc bitcoin cash bch toncoin ton dai avalanche avax shiba inu shib binance usd busd algorand algo hex cryptocurrency crypto\\
    & \\
    Domain keywords & kf event musk elon give coin shiba drop double get doge kefu vitalik claim binance hoskinson free charles star garling\\
    & \\
    HTML keywords & giveaway participate send address rules crypto bonus immediately hurry \\
    \bottomrule
\end{tabularx}
\caption{Keywords used to find relevant streams.}
\label{tab:search-keywords}
\end{table*}

\begin{figure*}[t]
\centering
\includegraphics[width=\textwidth]{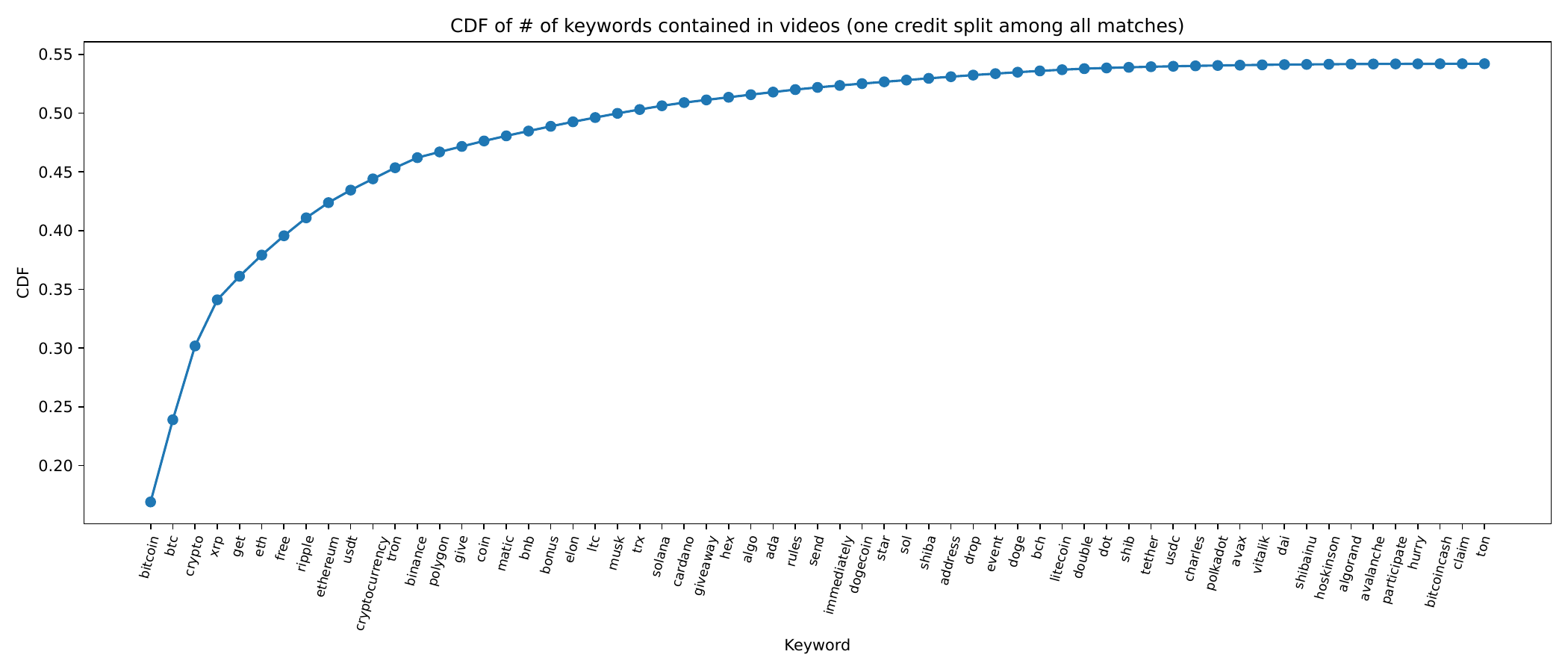}
\caption{Effectiveness of keywords}
\label{fig:keyword-effectiveness}
\end{figure*}

\section{Ethics}
Our work relies largely on public APIs (e.g., those provided by YouTube and Twitch) and public data (e.g., blockchain data and public tweets). We do not seek to deanonymize individuals involved in the transactions we explore, and aim to reduce this risk by obtaining only the category of cryptocurrency services as attribution tags, rather than the name of individual services.  Finally, we conducted all of our analysis retrospectively, which means we did not have the opportunity to report the identified scams as they were being promoted and thus potentially help their victims. 

\section{Stream Retrieval Pilot Study}
In this section, we provide more details on our pilot study conducted July 1-14, 2023. We start by discussing our attempt to identify scam livestreams on Twitch (Section~\ref{subsec:twitch_details}). Our search yielded no results, and thus we excluded Twitch from future experiments. 
We then describe how the various keywords we used for searching for
livestreams on YouTube incrementally contributed to the search results.
\label{app:pilot-study}

\subsection{Twitch details}
\label{subsec:twitch_details}
In this section, we discuss our methodology for identifying potential scam livestreams on Twitch.

\paragraph{Identifying relevant streams}
The Twitch API 
allows callers to retrieve all livestreams at a given point in time. We retrieved the list of all livestreams every 30 minutes, which is roughly the amount of time we needed to record and process the  streams in that batch. 

After retrieving all streams
we identified the ones that are the most worth recording; i.e., that seem most likely to be promoting giveaway scams. (The YouTube API already performs filtering, as it returns only streams related to the provided keywords.)
To filter out irrelevant streams on Twitch, we used their title, tags, and category.  We started by marking a stream as relevant if its tags or title contain any relevant keywords. We derived our set of relevant keywords from the keywords used by CryptoScamTracker, but removed 16 non-crypto keywords (e.g., ``event'' and ``give'') as they introduced too many irrelevant results. After filtering out streams that are not relevant, we then further removed any streams that are games using the category of the stream. This step effectively limits the number of streams to around 250.


\paragraph{Recording duration in pilot study}
Twitch
randomly inserts a 15-second advertisement clip before the actual stream content. To account for this issue, we recorded every Twitch stream for 20 seconds. 
For chat messages, Twitch's API does not provide history for chat messages. As a result, we recorded the Twitch chat for the entire duration of the stream.

\paragraph{Stream and chat recording}
We recorded streams and chats at our maximum capacity: sampling a stream for 20 seconds every 30 minutes (the minimum time needed to record and process a batch of streams) and recording the chat messages of a stream while it is live. We extracted any URL embedded in the stream as a QR code or in the chat messages. 

We ran this experiment for non-gaming streams that we identified in the filtering module as well as for 250 gaming streams that contained the most keywords. We did not find any scam livestreams during our 14-day pilot study, and thus excluded Twitch from further analysis.

\subsection{Keyword details}
\label{subsec:effectiveness_of_search_keywords}

We examine the incremental contribution of each search keyword (Table~\ref{tab:search-keywords}) to the set of livestreams returned from the YouTube API by computing the number of YouTube streams (over the 14-day pilot study) that contain search keywords in their metadata (title and description). If one stream contains multiple keywords, we give credit evenly among all keywords. Figure~\ref{fig:keyword-effectiveness} shows the result. Overall, around 55\% of the streams contained at least one of the search keywords, and the top 20 keywords account for 90\% of those streams. Among the streams that do not contain any search keyword, roughly 50\% are not in English. Sampling English streams that do not contain any keyword, we find that they are either on crypto-adjacent topics (e.g., trading) or completely irrelevant topics (e.g., a live street camera).

\end{document}